# 5G Configured Grant Scheduling for 5G-TSN Integration for the Support of Industry 4.0

Ana Larrañaga[1], M. Carmen Lucas-Estañ[2], Imanol Martinez[1], Javier Gozalvez[2]
[1]HW and Communication Systems Area, Ikerlan Technology Research Centre, Mondragón 20500, Spain
[2]UWICORE Laboratory, Universidad Miguel Hernández de Elche (UMH), Elche 03202, Spain
{ana.larranaga, imartinez}@ikerlan.es, {m.lucas, j.gozalvez}@umh.es

*Abstract*— **Factories are evolving towards digitalized data-based ecosystems under the paradigm of the Industry 4.0 where new industrial services allow the implementation of more robust, resilient and customized manufacturing systems. Such services (e.g., digital twins, extended reality or cooperative robots) will require highly reliable and deterministic communication networks capable of supporting stringent latency and reliability requirements. 5G networks and their future evolution have the necessary capabilities to meet these requirements. However, the use of 5G in industrial environments requires its effective and efficient integration with Time Sensitive Networking (TSN), which is becoming the standard wired technology for Industry 4.0 environments. TSN provides unprecedented deterministic service levels with perfectly bounded latencies. The integration of the industrial 5G and TSN networks will be key to support the flexibility and determinism demanded by the Industry 4.0 paradigm. A critical aspect to achieve this integration is the coordination of the schedulers of both networks. TSN has information about the capabilities of the 5G-TSN integrated network, and it is in charge of deciding the path and scheduling for each TSN traffic flow. The scheduling in 5G must be done according to the scheduling decisions and information provided by TSN to guarantee the end-to-end latency requirements of TSN traffic. In this context, this paper proposes a novel Configured Grant scheduling scheme for 5G integrated into a TSN network that aims to meet the latency requirements of the different TSN flows. The proposed scheme exploits the information provided by TSN about the characteristics of the TSN traffic to coordinate its decision with the scheduling of TSN. This study demonstrates that the proposed scheduling scheme considerably increases the number of TSN flows that can be satisfactorily served in the integrated 5G-TSN network compared with a commonly used Configured Grant (CG) scheduling scheme.**

***Keywords—5G; 5G-TSN integration; Deterministic; Low Latency; TSN; Configured Grant; scheduling; Industry 4.0.***

## I. INTRODUCTION

Factories are evolving towards digitalized data-based ecosystems where new industrial services, such as digital twins, extended reality (XR) or cooperative robots, emerge and allow the implementation of more robust and resilient manufacturing systems. Industrial environment and processes will be accurately monitored in real-time thanks to Industrial IoT (Industrial IoT or IIoT) networks that will connect machines, mobile robots, sensors, and user terminals, allowing data to be collected, analyzed, and distributed instantly [1]. Factories of the future will then require highly reliable communication networks capable of supporting the stringent latency, bandwidth, and reliability requirements of demanding industrial applications. In addition, the communication networks that will support Industry 4.0 must be able to adapt their operation to the changing connectivity and data transfer needs of industrial production systems. 5$^{th}$ Generation (5G) networks and their future evolution have the necessary capabilities to meet these requirements. In fact, 5G technology is considered a key catalyst for the digital transformation of the industry.

The use of 5G in industrial environments requires its effective and efficient integration with industrial wired networks currently implemented in factories to meet the strict reliability and resilience requirements of industrial applications. Time-Sensitive Networking (TSN) is becoming the standard wired technology for Industry 4.0 environments. TSN is based on Ethernet and provides unprecedented deterministic service levels with perfectly bounded latencies. However, TSN does not have the flexibility and reconfiguration capabilities required by the factories of the future. On the other hand, 5G is highly reconfigurable and flexible but cannot efficiently and scalably support deterministic services. Both networks, therefore, have complementary capabilities and the integration of the industrial 5G and TSN networks is key to support the Industry 4.0 paradigm.

3GPP standards already define the framework for the efficient integration of TSN and 5G networks. The 5G network is integrated into the TSN network as a logical TSN bridge of the TSN network (see Fig. 1). A TSN bridge is an Ethernet switch that receives and transmits TSN frames based on a scheduling. In this integrated network, the communication path between two end-devices (for example, a sensor and an actuator) is established through one or more TSN bridges and the 5G logical bridge. In this context, a critical aspect to meet the end-to-end latency requirements of the industrial applications is the coordination of the scheduling of both networks. In the integrated 5G-TSN network, the Central Network Configuration (CNC) has information about the whole network topology and capabilities of the bridges. Based on this information, the CNC decides the communication path and scheduling for each TSN traffic flow. The scheduling decision establishes the arrival and departure time of the packets of each TSN flow in each (TSN and 5G logical) bridge. The scheduling of the packets of the TSN flows in the 5G network has to be done in a way that these arrival and departure times calculated by TSN are satisfied.

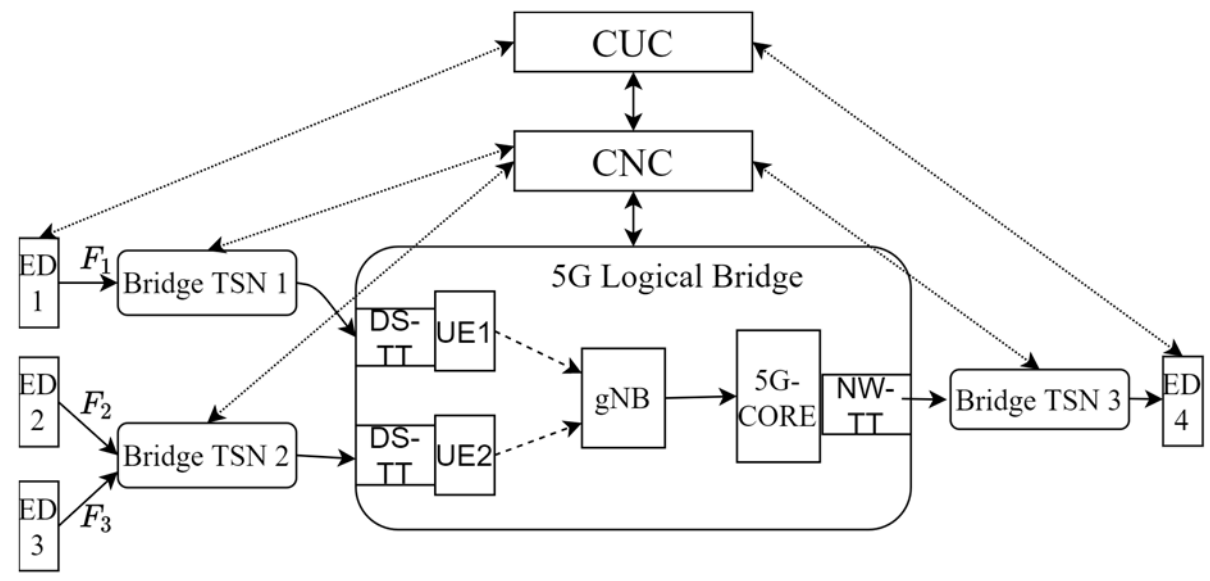

Fig. 1. 5G-TSN integration model.

Some current researcher work have studied scheduling schemes for 5G networks integrated into TSN networks as a logical bridge. For example, [2] proposed a dynamic scheduling scheme for deterministic traffic. The proposed scheme is based on optimization techniques and allocates resources dynamically for each TSN packet to improve radio resource efficiency while guaranteeing the latency requirements of the TSN traffic. With dynamic scheduling, the end users need to request resources and/or receive a grant before transmitting a packet. This exchange of signaling can increase the latency of communication. Configured Grant (CG) in uplink (UL) and Semi-Persistent Scheduling (SPS) in downlink (DL) allocate radio resources periodically for each TSN flow based on the periodicity of the packets. Configured Grant or Semi-Persistent Scheduling are the most proper scheduling (in UL and DL, respectively) for this type of traffic based on its low latency requirements and periodicity. An SPS scheme has been studied in [3] for DL Time Sensitive Communications (TSC) flows. The authors in [3] proposed the adaptation of the modulation and coding scheme (MCS) used for the transmission of each packet based on past channel conditions. [4] proposed a predictive multi-priority scheduling mechanism based on SPS for the 5G-TSN network where periodic and aperiodic data is considered. According to [5], CG or SPS are the most proper scheduling (in UL and DL, respectively) for TSC traffic demanding low latency requirements. The allocation of periodic resources to TSN flows with different periodicity may result in conflicts: the resources allocated to different TSN flows may overlap after several periods. This is illustrated in Fig. 2. The example in Fig. 2 shows two TSN flows, $F_1$ and $F_2$, with periodicity 3 and 5 respectively. Although the radio resources allocated for the transmission of the first packet of each TSN flow are different, some of the resources allocated for the transmission of the third and fifth packets of F1 and F2 respectively overlap in frequency and time due to the different packet periodicities of F1 and F2. This is a critical issue that, to the best of the author's knowledge, has not been previously addressed. In this context, this paper progresses the state of the art by proposing a novel scheduling scheme for 5G integrated into TSN networks that addresses this problem. The main objective of the proposed scheduling scheme is to meet the latency requirements of the different TSN flows based on the arrival and departure times calculated by the TSN network. To this end, the proposed scheduling scheme uses the information provided by the TSN network about the characteristics of the TSN flows (packet sizes, periodicity, etc). The proposed scheduling scheme configures several grants for each TSN flow in order to avoid radio resource allocation conflicts among different TSN flows. This study demonstrates that the proposed scheduling scheme considerably increases the number of TSN flows that can be satisfactorily served in the integrated TSN-5G network.

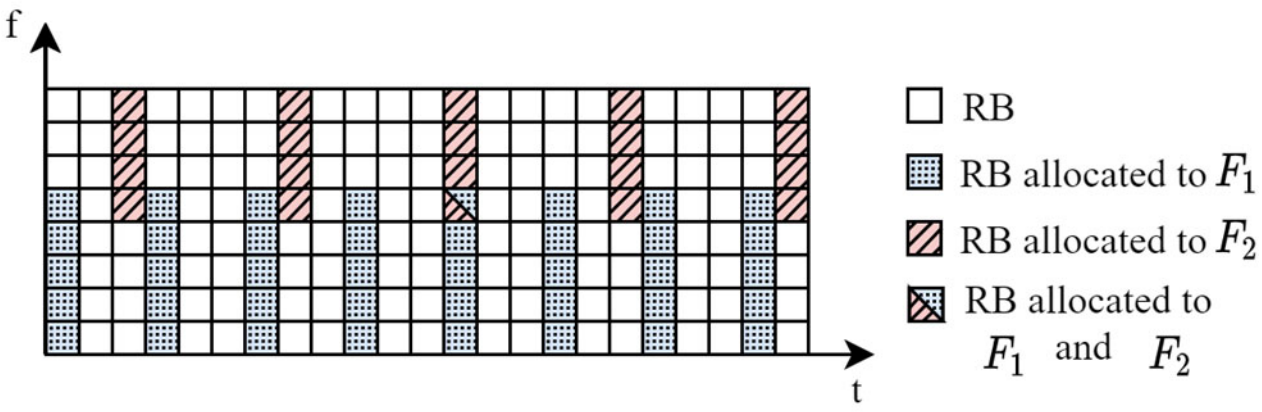

Fig. 2. Example of a radio resource allocation conflict where the same radio resources are allocated to more than one TSN flow when radio resources are allocated considering the packet periodicity.

The paper is organized as follows. Section II presents the current framework for TSN and 5G networks integration. Section III presents the proposed 5G scheduler for 5G integrated into a TSN network. Section IV presents the evaluated scenario and Section V presents the performance achieved with the proposed scheme. Section VI concludes the paper.

## II. 5G-TSN Integration

TSN is a set of open standards developed by the IEEE 802.1. TSN has been designed to provide communications with a very high level of determinism: TSN guarantees the delivery of the data in a guaranteed time window [6]. This is achieved thanks to the introduction of several technical features, such as strict time synchronization (IEEE 802.1AS) and a traffic scheduler (IEEE 802.1Qbv) that reserves specific time slots for the transmission of high-priority traffic. 5G networks have the capacity to support low latency and highly reliable communications. 5G can complement TSN to provide the level of determinism and flexibility demanded by Industry 4.0. 3GPP standards define the framework for the integration of 5G in TSN networks. A TSN network is integrated by end devices (ED), which are the source and destination of TSN flows, and bridges that are interconnected using standard Ethernet links. In the integration model defined by the 3GPP, 5G appears as a (logical) bridge of the TSN network, as shown in Fig. 1 5G includes TSN translator (TT) functionality at the edges of the network: at the users (UEs) referred to device side TT or DS-TT and at the 5G Core Network referred to network side TT or NS-TT. The TTs are the interconnection points between 5G and TSN. The TTs allow 5G to inform the TSN network about its capabilities and current status, as well as to receive, interpret and apply TSN configuration commands from the TSN network. The TTs provide TSN bridge ingress and egress port operations. For example, TTs are responsible for holding and forwarding the TSN flows to minimize latency variation.

The integrated 5G-TSN network considers a centralized management model. The TSN network incorporates the Centralized User Configuration (CUC) and CNC functions. The CUC receives information about the communication requirements for each TSN flow from the end devices. The CUC informs the CNC about these requests. The CNC collects information about the capabilities and current status of all the (TSN and 5G logical) bridges in the network [7][8]. For example, bridges inform about the minimum and maximum supported bridge delays, which are the minimum and maximum times respectively that a packet needs to be forwarded from an ingress port to an egress port of the bridge. TSN bridges also inform of the propagation delay or link delay, which is the time a packet takes to travel through the links that interconnect the nodes (end devices and bridges) from the specified port of the station to the neighboring port on a different station. The CNC then decides the path and schedule of all TSN flows using IEEE 802.1Q [9]. The CNC finally configures the (TSN and 5G logical) bridges according to the scheduling decision in order to guarantee the end-to-end requirements of the TSN flows. The CNC applies the time-aware shaper (TAS) defined in IEEE 802.1Qbv that establishes the arrival and departure time of each TSN flow at the ingress and egress ports of each TSN and 5G logical bridge, respectively.

5G translates the information about TSN QoS (Quality of Service) requirements received from the CNC to a 5G QoS Identifier (5QI) for each TSN flow. A 5QI defines a set of 5G QoS characteristics that describe the packet forwarding treatment that a QoS flow receives in the 5G network (between the UE and the last node in the 5G core network). A 5QI includes the priority level of a flow for the scheduling of resources, the packet error rate (PER), the maximum data burst volume (MDBV), and the packet delay budget (PDB), which indicates the maximum latency a packet can experience in the 5G network, among other QoS characteristics. 3GPP standards do not specify how TSN QoS requirements are mapped to 5G QoS requirements. Some proposals are found in [10], [11], and [12]. In addition to the 5QI, 5G determines Time Sensitive Communication Assistance Information (TSCAI). The TSCAI describes the TSN traffic characteristics: burst or packet arrival time with reference to the ingress port, periodicity, flow direction (uplink or downlink), and survival time[1] [8]. The TSCAI and the 5QI may be used by the gNB (new generation Node B) to decide the scheduling of radio resources for the TSN traffic in order to meet the end-to-end requirements of the TSN flows.

## III. 5G CG Scheduling for the Support of TSN Traffic

This work proposes a scheduling scheme for a 5G network integrated into a TSN network to support the stringent communication requirements of industrial applications. The main objective of the proposed scheduling scheme is to guarantee the 5G latency requirements for each TSN flow and ensure that packets are received at the egress port of the 5G logical bridge before the departure time indicated by the TSN network. To this end, the proposed scheduling scheme uses the information provided by the TSN network (5QI and TSCAI) to decide the radio resource allocation solution that better satisfies the latency requirements of all TSN flows. A radio resource is composed of one Resource Block (RB) of 12 subcarriers in the frequency domain and one Orthogonal Frequency Division Multiplexing (OFDM) symbol in time domain. The scheduling proposal is based on CG, and it allocates radio resources periodically for each TSN flow. The proposed scheduling scheme configures several grants for each TSN flow, if needed, to ensure that radio resources are not assigned to more than one TSN flow simultaneously.

Let us consider that $N_F$ TSN flows must be transmitted through the 5G virtual bridge toward the end devices. For each TSN flow $F_i$ (with $i$=1,…, $N_F$), packets of size $s_i$ arrive with a periodicity $p_i$; $s_i$ is given by the MDBV of the 5QI and $p_i$ in the TSCAI. $s_i$ and $p_i$ can be different for each TSN flow. We consider that all the TSN traffic that arrives to the 5G network must be transmitted in UL (flow direction is represented by $f_i$). Each flow demands $d_i$ radio resources calculated as a function of $s_i$ and the MCS to use for the packet transmission. Packets in a TSN flow $F_i$ are referred to as $pkt_{i,j}$ where $j$ indicates the number of the packet in the flow. The arrival and departure time instants of a packet $pkt_{i,j}$ are given by $A_{i,j}=A_i+j\cdot p_i$ and $D_{i,j}=A_{i,j}+l_{5G_i}$, respectively, where $A_i$ is the burst arrival time indicated in the TSCAI and $l_{5G_i}$ is the latency requirement for packets in the TSN flow $F_i$ indicated by the PDB of the 5QI. A TSN flow $F_i$ is then defined as $F_i=\{A_i, l_{5G_i}, d_i, p_i, f_i\}$. The proposed scheduling scheme exploits this information to decide the radio resources that must be allocated to each TSN flow.

The TSN flows supported by the 5G network transmit periodic packets. Although the periodicity $p_i$ of each TSN flow can be different, it is possible to find a pattern of packet arrival considering all TSN flows that repeat periodically. The periodicity of this packet arrival pattern is referred to as hyperperiod or *HP*. Fig. 3 shows an example with three TSN flows with different periodicities (30, 45 and 90 ms). It is possible to see in that there is a packet arrival pattern that repeats every 90 ms when considering the set of all TSN flows. The scheduling scheme identifies *HP* and allocates radio resources for each packet within *HP*. The allocated radio resources for each TSN flow are pre-assigned periodically with a periodicity *HP*.

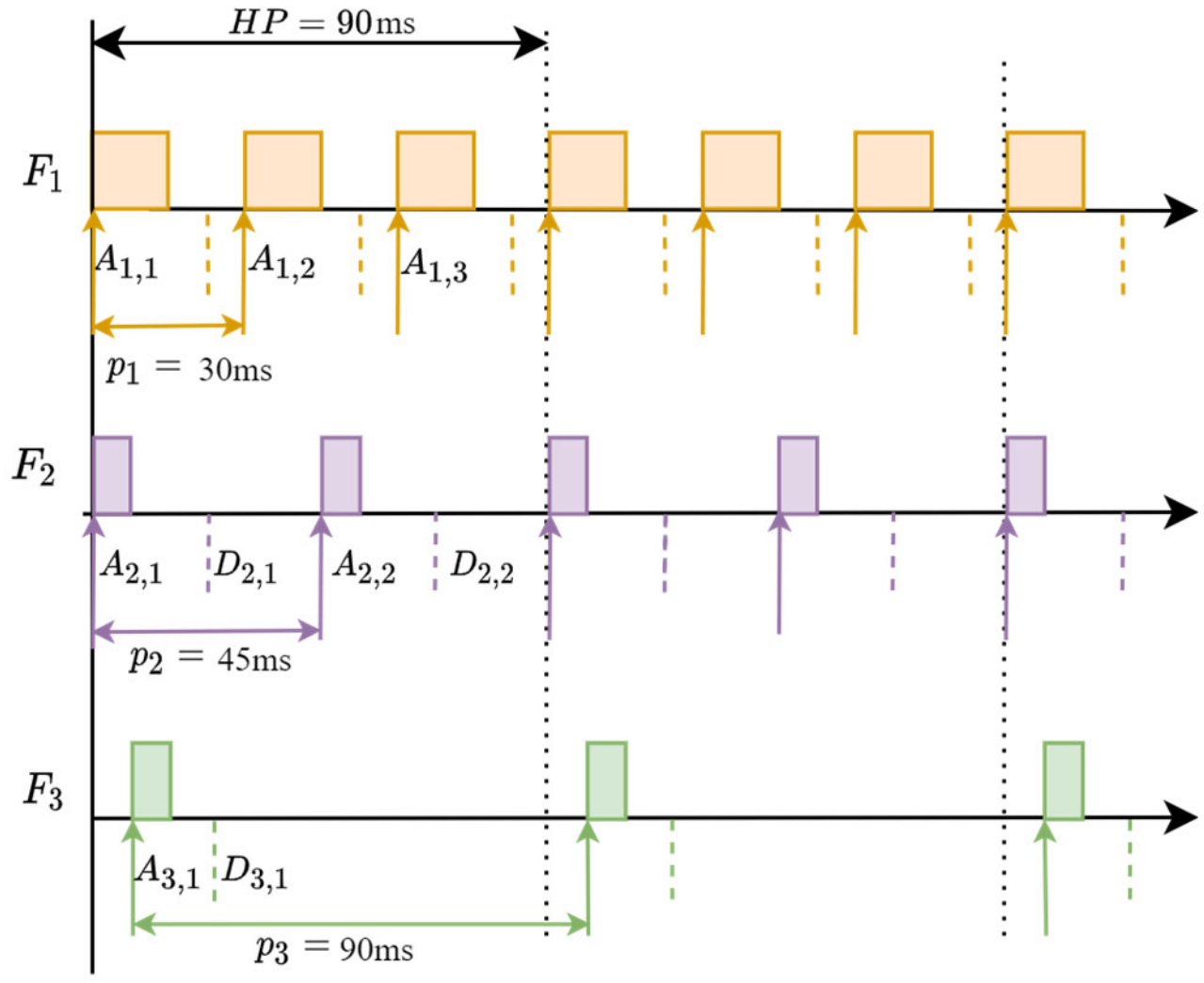

Fig. 3. Hiperperiod for an example case with 3 TSN flows with different periodicities.

[1] Survival time is the time period an application can survive without receiving any burst or packet as described in [24].

The operation of the proposed scheduling scheme is presented in Algorithm I and Algorithm II. The scheduling scheme first calculates *HP*. *HP* is calculated as the least common multiple (lcm) of the periodicity of the packets transmitted in each TSN flow, i.e., $HP = \text{lcm}(p_i)$, with $i \in [1, N_F]$ (line 2 in Algorithm I). The packets of all the TSN flows within the time window given by *HP* are divided into different groups or blocks. Each block contains the packets whose transmission could overlap in time considering their arrival and departure times. Two packets $pkt_{i,j}$ and $pkt_{m,n}$ overlap if $A_{i,j} \leq A_{m,n} \leq D_{i,j}$ or $A_{i,j} \leq D_{m,n} \leq D_{i,j}$ (see Algorithm II). In the example of Fig. 3, two blocks are created: the first block $B_1$ contains $pkt_{1,1}$, $pkt_{2,1}$ and $pkt_{3,1}$, while the second block $B_2$ contains $pkt_{1,2}$, $pkt_{2,2}$ and $pkt_{1,3}$. The proposed scheme addresses the radio resource allocation process separately for each block $B_z$, with $z$=1,2,... First, the scheduling scheme calculates the number of symbols $d_i^S$ and RBs $d_i^R$ that should be allocated for the transmission of each packet to satisfy $d_i$. To this end, the scheduling scheme is based on the Sym-OFDMA (Orthogonal Frequency Division Multiple Access) scheduling policy presented in [13]. Sym-OFDMA has been selected because it reduces the transmission latency compared to other scheduling policies. Following Sym-OFDMA, the scheduling scheme tries to minimize the number of symbols assigned for the transmission of each packet. In this context, the scheduling scheme will allocate $d_i^R = d_i$ RBs in one symbol ($d_i^S$=1) when $d_i$ is lower than the number of RBs available in the bandwidth ($R_{BW}$), i.e. $d_i \leq R_{BW}$,. If $d_i > R_{BW}$, the scheduling scheme will allocate $d_i^R = R_{BW}$ RBs in $d_i^S = \lceil d_i / R_{BW} \rceil$ symbols. When $d_i^S$ and $d_i^R$ are known, the scheduling scheme starts an iterative process. At each iteration, the scheduling scheme tries to find a radio resource allocation solution that satisfies the latency and departure times for all packets. The scheduling scheme starts a new iteration if it cannot find adequate radio resources to satisfy the departure time for a packet. The iterative process will finalize when one of the two following conditions are met: all packets in $B_z$ have allocated resources, or it is not possible to find a feasible solution to the problem. At each iteration, the proposed scheduling scheme serves packets considering their latency requirements $l_{5G_i}$: packets with lower $l_{5G_i}$ are served first (line 8 in Algorithm I). The scheduling scheme allocates

ALGORITHM I: *SCHEDULING*

1. Input: $F_i$ ∀ $i \in [1, N_F]$, $t_{sym}$
2. HP = $\text{lcm}\{p_i\}$ with $i \in [1, N_F]$
3. Create packet blocks $B_z$, with $z$=1, 2, … (Algorithm II)
4. Initialize $BF_z \neq \emptyset$
5. **For** $z$=1 to $N_B$
6. Calculates $d_i^S$ and $d_i^R$ for all packets in $B_z$
7. $N_{pkt}$ = number of packets in $B_z$
8. $pkt_{I,j}$ = packet with lower $l_{5G}$ in $B_z$
9. $B_z = B_z - \{pkt_{i,j}\}$
10. $iter$=0, $BF_z = \emptyset$
11. **For** $iter$ = 0 to ∞
12. **While** $B_z \neq \emptyset$
13. $s$ = first OFDM symbol after $A_{i,j}+t_{UE,tx}$
14. **While** $s + d_i^S$-1 <= $D_{i,j}$
15. **If** there are $d_i^R$ unallocated RBs in symbols $s$ until $s+ d_i^S$-1
16. $pkt_{i,j}$ receives $d_i^R$ RBs in symbols $s$ until $s+ d_i^S$-1
17. First allocated symbol: $s_{i,j} = s$
18. **If** $s + d_i^S$-1 == $D_{i,j}$
19. $BF_z = BF_z + \{pkt_{i,j}\}$
20. **EndIf**
21. **Else**
22. $s$++
23. **EndIf**
24. **EndWhile**
25. **If** $s + d_i^S$-1 > $D_{i,j}$
26. $BF_z = BF_z + \{pkt_{i,j}\}$
27. **If** $BF_z + B_z == N_{pkt}$
28. Goto 41
29. **Else**
30. Reinitiate $B_z$
31. $B_z = B_z$ - $BF_z$
32. Free radio resources allocated to packets in $B_z$
33. Goto line 11
34. **EndIf**
35. **Else**
36. $pkt_{I,j}$ = packet with lower $l_{5G}$ in $B_z$
37. $B_z = B_z - \{pkt_{i,j}\}$
38. **EndIf**
39. **EndWhile**
40. **EndFor**
41. **End For**

ALGORITHM II: *BLOCK CREATION*

1. Input: $F_i$ ∀ $i \in [1, N_F]$
2. $I=\{A_{i,j}\}$ $\forall i \in [1, N_F]$, $\forall j \in [0, HP/p_i]$
3. $z$=1, $B_z=\emptyset$
4. **While 1**
5. $\{i,j\} = \arg_{m,n}\left\{\min_{m,n}\{I\}\right\}$ $\forall m \in [1, N_F]$, $\forall n \in [1, HP/p_m]$
6. $I = I - \{A_{i,j}\}$
7. **If** $B_z == \emptyset$
8. $D_{aux} = D_{i,j}$, $A_{aux} = A_{i,j}$
9. $B_z = \{pkt_{i,j}\}$
10. **Else**
11. **If** $A_{i,j} <= D_{aux}$
12. $B_z = B_z \cup \{pkt_{i,j}\}$
13. **If** $D_{i,j} > D_{aux}$
14. $D_{aux} = D_{i,j}$
15. **End If**
16. **Else**
17. $z$=$z$+1, $B_z=\emptyset$
18. Go to 7
19. **End If**
20. **End If**
21. **If** $I==\emptyset$
22. Go to 24
23. **End If**
24. **End While**

to each packet $pkt_{i,j}$ $d_i^R$ RBs in the first $d_i^S$ consecutive symbols with at least $d_i^R$ unallocated RBs after the packet is generated ($A_{i,j}$+$t_{UE,tx}$, where $t_{UE,tx}$ represents the processing time in the transmitter to generate the packet) and before the departure time $D_{i,j}$ (lines 15-21 in Algorithm I). If a packet receives radio resources just before its departure time $D_{i,j}$, the packet will maintain the allocated radio resources although a new iteration is performed. To this end, it is included in a set $BF_z$. $BF_z$ includes the packets that cannot change their allocated radio resources in potential next iterations (see lines 18-20 in Algorithm I). If it is not possible to find the required RBs and OFDM symbols for a packet, the scheduling scheme will start a new iteration. Before that, $B_z$ is reinitialized but excluding packets in $BF_z$ (lines 25-35 in Algorithm I). The first packet served in the new iteration will be the packet that was not possible to assign radio resources in the previous iteration. This packet is also included in $BF_z$ (line 26 in Algorithm I). The scheduling process finalizes when all packets have allocated resources or when it is not possible to find a feasible solution to the problem (lines 27-29).

## IV. Evaluation Scenario

We consider an evaluation scenario where a 5G network is integrated into a TSN network as a logical bridge. The integrated 5G-TSN network provides connectivity to an industrial plant where there is implemented a closed-loop supervisory controller application [14]. This application integrates a Programmable Logic Controller (PLC) which receives monitoring data from sensors (S1, S2, …, $S_{N_F}$) that creates a total of $N_F$ TSN flows. Based on the received data, the PLC sends a command to an actuator (A). Fig. 4 shows the evaluated scenario. The PLC, sensors, and actuator are the static end devices that exchange data in the network. The end devices are connected through several bridges, including the 5G logical bridge. The evaluation of the proposed scheduler has been carried out using Matlab.

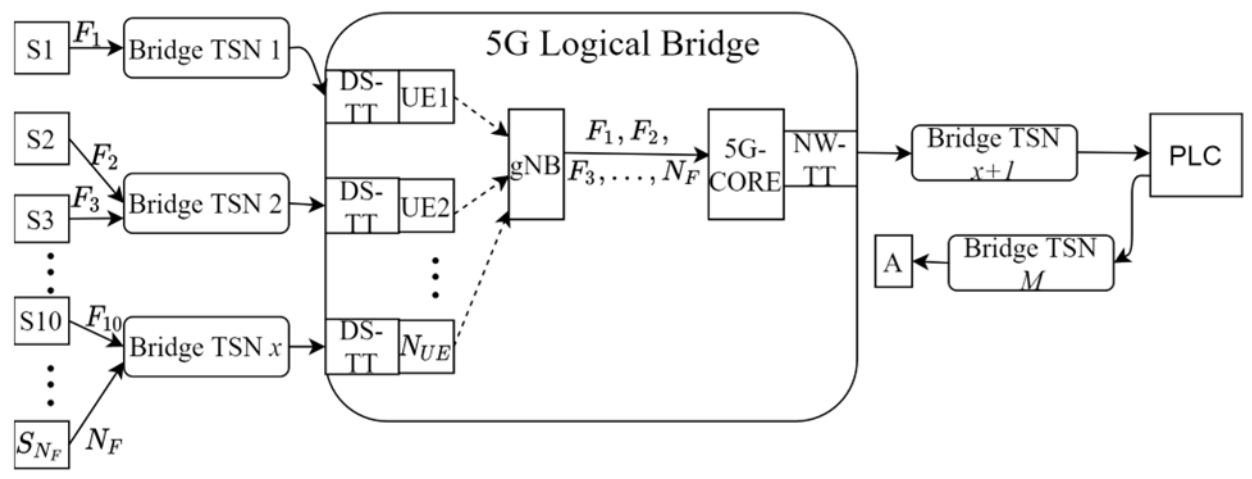

Fig. 4 . Evaluated scenario.

The sensors send the sensed data to the PLC. This results in $N_F$ TSN flows between the sensors and the PLC that are transmitted through the 5G logical bridge (we consider $N_F$ equal to 30). Each TSN flow $F_i$ is defined by the size of the transmitted packets ($s_i$), the periodicity of the packets ($p_i$) and the demanded end-to-end latency ($l_{e2e_i}$) between the sensor and the PLC. Based on [14], $l_{e2e_i}$ can take values between 4ms and 20ms. We consider that the periodicity $p_i$ is equal to the $l_{e2e_i}$ requirement (this means that each packet needs to be received before the next packet is generated). TSN flows randomly select the periodicity $p_i$ between $p_1$, $p_2$ or $p_3$, and $p_j$, where j ϵ {1,2,3}, is randomly selected between 4 and 20 ms. The packet size $s_i$ for each TSN flow $F_i$ can take a value between 40 and 250 bytes. The characteristics and requirements that need to be satisfied by the 5G logical bridge of a TSN flow is indicated to the 5G network using the TSCAI and 5QI information. For each TSN flow, the TSN network informs the 5G network about the following information: the periodicity $p_i$ and packet size $s_i$ of the packets, the flow direction $f_i$, the arrival time $A_i$ at the ingress port of the 5G logical bridge, and the maximum latency requirement of the TSN flow inside the 5G logical bridge $l_{5G_i}$. In the evaluation scenario, all TSN flows in the 5G network are transmitted in uplink, i.e., from the UEs to the gNB. The arrival time $A_i$ and the 5G latency requirement is calculated by the CNC that has information about the TSN network topology and is in charge of the scheduling of the TSN flows in the TSN network to satisfy the end-to-end latency requirements. The arrival time $A_i$ of the first packet of a TSN flow is equal to the latency that the packet experiences from the moment it is generated in the sensor until it is received in the 5G logical bridge. Then, it depends on the application processing time in the sensor $l_{sensor_i}$, the number and the processing time of TSN bridges that there are in the path between the sensor that generates the data and the 5G logical bridge $l_{TSN_{b,i}}$ with $b \in [a,..., x]$, the propagation time it takes a packet to travel through the links that interconnect the nodes (end devices and bridges) $l_{links_i}$, and the processing time $t_{DS\text{-}TT}$ at the DS-TT translator. The $A_i$ is given by the following expression:

$$A_i = l_{sensor_i} + l_{links_i} + l_{TSN_{a,i}} + \ldots + l_{TSN_{x,i}} + t_{DS\text{-}TT} \qquad (1)$$

$l_{TSN_{b,i}}$ represents the time that packets of a TSN flow $i$ spend in a TSN bridge $b$. $l_{TSN_{b,i}}$ is null if TSN bridge $b$ is not in the path between the sensor that generates the data and the 5G logical bridge. Otherwise, $l_{TSN_{b,i}}$ considers processing and queuing delays in the TSN bridge (independent delay $t_{indep_{b,i}}$) that vary depending on the traffic load [8], and the transmission delay that depends on the packet size and the bit rate used by the TSN bridge to transmit the data to the next node in the path (dependent delay $t_{dep_{b,i}}$). To calculate the arrival time, we consider that all sensors (30 sensors) connect to a TSN bridge that connects to the 5G logical bridge, and all sensors transmit packets of 250 bytes. Due to short distances, $l_{links_i}$ is considered negligible. We consider DS-TT and NW-TT processing times equal to 0.05 ms and 0, respectively [15]. We assume a data bit rate equal to 100 Mbps, and the minimum independent delay $t_{indep_{b,i}}$ is equal to the TSN bridge processing delay, which is 1.5 µs as presented in [16]. Then, for the minimum arrival time, we assume that a flow is transmitted through the TSN bridge directly to the 5G logical bridge as soon as it arrives to the ingress port of the TSN bridge. We have considered that the sensor processing time is 3 µs [17], then, the minimum arrival time to 5G logical bridge for a packet of 250 bytes is $3\mu s + \left(\frac{250 \cdot 8\text{bits}}{100\text{Mbps}} + 1.5\mu s\right) + 50\mu s \cong 75\mu s$. In the case of 700µs, which represents the maximum arrival time, we consider the last flow to be transmitted from TSN bridge to the 5G logical bridge arrives at $3\mu s + 30 \cdot \left(\frac{250 \cdot 8 bits}{100 Mbps} + 1.5\mu s\right) + 50\mu s \cong 700\mu s$ to the ingress port of 5G logical bridge. This results in arrival times between 75µs and 700 µs.

We calculate the latency that must be satisfied in the 5G network for each TSN flow $F_i$ as:

$$l_{5G_i} \leq l_{e2e_i} - A_i - l_{links_i} - \sum_{b=x+1}^{M} l_{TSN_{b,i}} - t_{NW\text{-}TT} - l_{PLC_i} \quad (2)$$

In (2), $l_{PLC_i}$ represents the application processing time in the PLC, $t_{NW\text{-}TT}$ is the processing time at the NW-TT translator, $l_{TSN_{b,i}}$ represents the time that packets of a TSN flow $i$ spend in a TSN bridge $b$ with $b \in [x+1,..., M]$, $M$ is the number of TSN bridges in the TSN flow and $l_{links_i}$ is the propagation time it takes a packet to travel through the links that interconnect the nodes from the 5G logical egress port until the PLC ingress port. To calculate $l_{5G_i}$, we assume $l_{PLC_i}$ equal to 1007 µs as we consider a scan time of analog input data and the execution time of Proportional Integral Derivative (PID) control in the PLC of 1ms [18] and 7µs [19], respectively. We have assumed that the scan and processing times are minimal. This data is an approximation since it will vary with the application.

The 5G network deployed in the industrial plant implements a single cell with a bandwidth of 20 MHz. We consider the use of a 30 kHz sub-carrier spacing (SCS) as recommended in [20] for industrial environments. The cell operates in TDD mode [21], and each slot reserves 2 OFDM symbols for the transmission of UL and DL control messages and 12 OFDM symbols for the transmission of UL data. The UEs are placed in locations that guarantee Line of Sight conditions with the gNB. Based on that, we consider that all packets are transmitted using the MCS 12 in MCS table 1 [22] which guarantees a good compromise between robustness and spectral efficiency (modulation order $Q_m$ equal to 4 and coding rate $R$ equal to 434/1024). The number of radio resources to transmit a packet of size $s_i$ can be then calculated as:

$$d_i = \left\lceil \frac{(tbs_i(s_i + s_{header}) + s_{CRC}) \cdot 8}{R \cdot Q_m \cdot N_{sc,RB}} \right\rceil \quad (3)$$

In (3), $s_{header}$ represents the size in bits of the IPv4 header added to the data packet, and $s_{CRC}$ is the size in bytes of the cyclic redundancy check (CRC) code added at the end of the packet. $tbs(s_i+s_{header})$ represents the smallest transport block size from the available values given in [23] that can be used to transmit $s_i+s_{header}$ bits, and $N_{sc,RB}$ is the number of subcarriers in a resource block ($N_{sc,RB}$=12).

## V. Performance Evaluation

This section evaluates the performance of the proposed scheduling scheme proposed for a 5G network integrated into a TSN network to provide connectivity in industrial scenarios. To this end, we compare its performance to that achieved with a commonly used CG scheduling scheme [13]. The reference scheme allocates radio resources periodically for each TSN flow $F_i$ with a periodicity $p_i$. Each TSN flow $F_i$ receives the number of radio resources necessary to transmit packets of size $s_i$ according to (3). For a fair comparison with the proposed scheduling scheme, the reference scheme serves TSN flows based on their 5G latency requirements: TSN flows with lower $l_{5G_i}$ values are served first. Since the TSN flows can have different periodicities, several TSN flows could receive the same radio resources after some periods.

Due to the limited availability of radio resources and the high resource demands, it can be difficult for scheduling schemes to find a valid solution that can satisfy all requests. In this context, we first analyze the capabilities of the proposed and reference scheduling schemes to achieve a solution that allocates the requested radio resources for all the TSN flows. Fig. 5 shows the percentage of evaluated scheduling problems for which the proposed and reference schemes satisfy the radio resource request for all the TSN flows. The results in Fig. 5 are shown for different numbers of TSN flows demanding resources in the 5G network (between 10 and 30). Fig. 5 (lines with circles) considers that the end-to-end latency ($l_{e2e_i}$) demanded by each TSN flow $F_i$ is equal to its packet periodicity $p_i$, and Fig. 5 (lines with starts) considers that all TSN flows demand a $l_{e2e_i}$ equal to 4 ms. The results in Fig. 5 show that when the latency requirements are more relaxed, both the proposal and the reference scheme obtained solutions that satisfy all the requests. When the latency requirements are more stringent ($l_{e2e_i}$=4 ms for all the TSN flows) and the number of TSN flows requesting resources increases to 25, the reference scheme fails to find solutions satisfying all the requests in 9% of the cases. In this scenario, the proposal achieves solutions satisfying all the requests in the 98% of the cases. When the number of TSN flows increases to 30, the proposal is able to satisfy all the requests in the 33% of the cases, while the reference scheme only in the 17%.

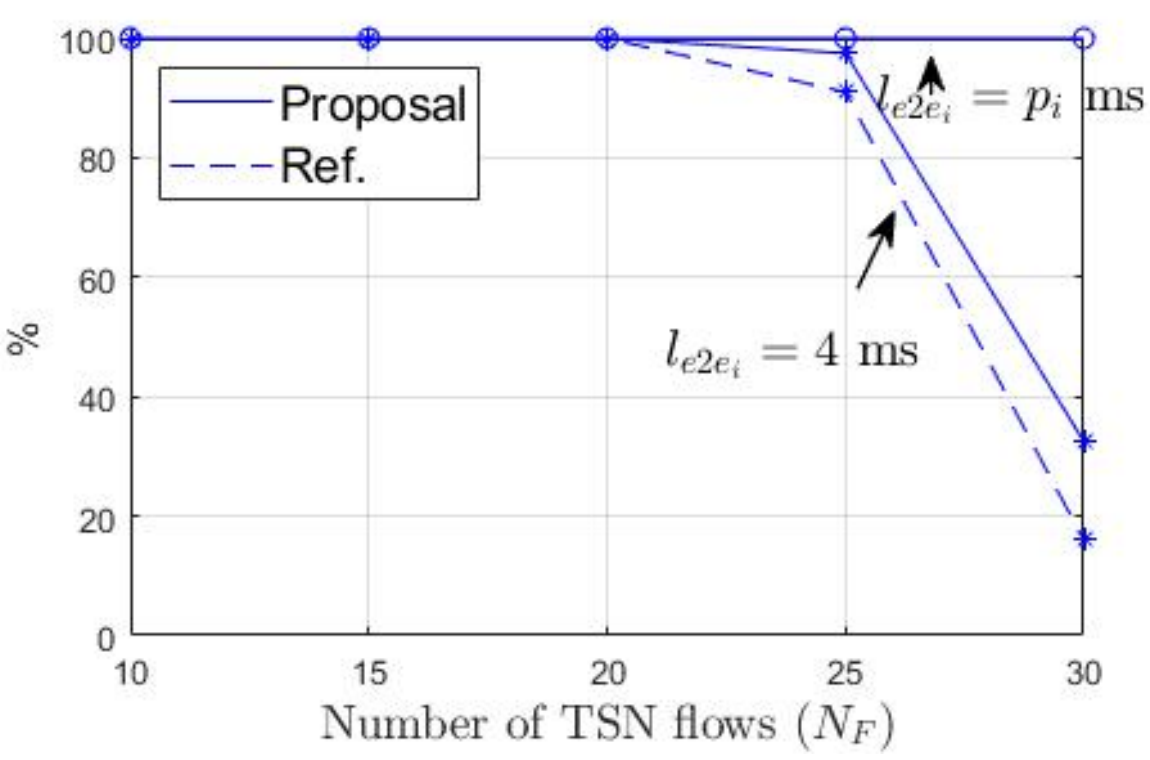

Fig. 5. Percentage of times the proposed and reference scheduling schemes satisfy the radio resource request for all packets of all TSN flows as a function of the number of TSN flows.

The solutions achieved with the reference scheme ensure that the radio resources allocated to the TSN flows do not overlap in time and frequency simultaneously for the first packet in each TSN flow. However, this is not guaranteed for the following packets in the flows, and several TSN flows can receive the same radio resources after several periods due to the different periodicities of each TSN flow (Fig. 2). This is a radio resource allocation conflict that can ultimately result in packet collisions. Fig. 6 shows the percentage of times that the reference scheme achieves a solution serving all requests, but some radio resources are allocated to more than one TSN flows. The results in Fig. 6 show that the number of times that the solution achieved with the reference scheme results in a radio resource allocation conflict increases with the number of TSN flows demanding resources. It is important to note that the percentage of times that there is a resource allocation conflict with the reference scheme is higher than 74% when the number of TSN flows demanding resources is equal to or higher than 15.

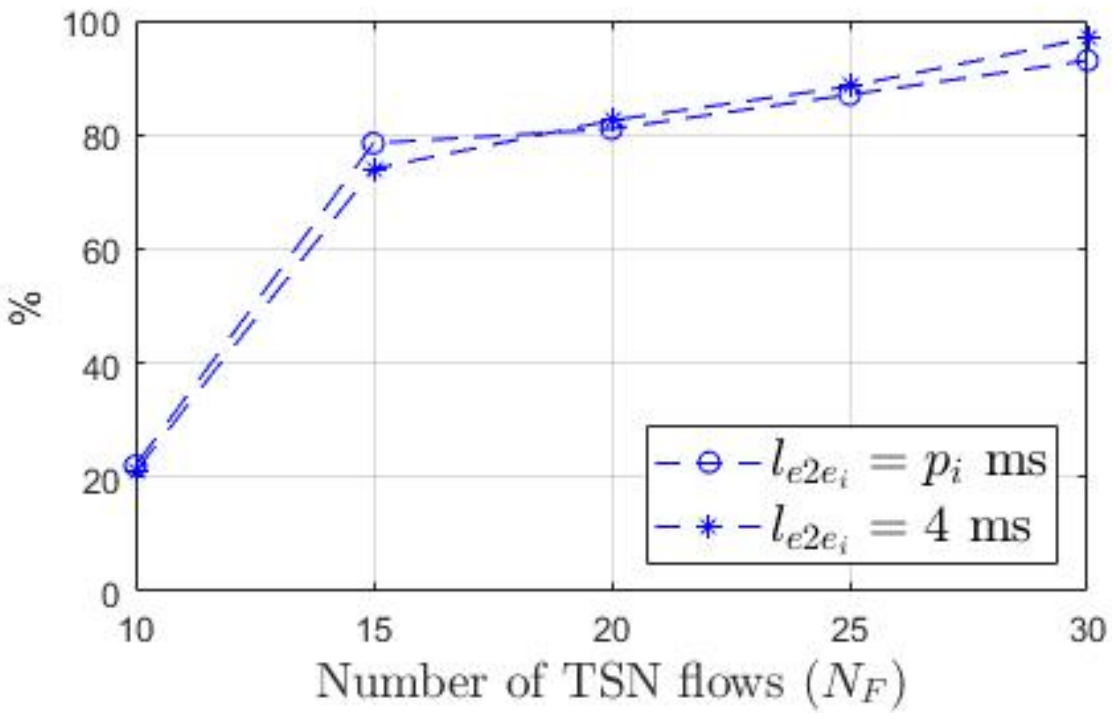

Fig. 6. Percentage of times that the reference scheme achieves a solution serving all requests, but some radio resources are allocated to more than one TSN flows as a function of the number of TSN flows.

Fig. 7 depicts the average of the latency and the average of the standard deviation experienced by the packet of each TSN flow when a different number of TSN flows are transmitted in the 5G network considering that the end-to-end latency ($l_{e2e_i}$) demanded by each TSN flow $F_i$ is equal to its packet periodicity $p_i$. The results show that the proposed scheduling scheme reduces the average latency experienced by the transmitted packets by more than 40% when the number of TSN flows is equal to or higher than 25. This is a result of the higher flexibility of the proposed scheduling scheme. With the reference scheme, a TSN flow receives resources periodically with the same periodicity that packets arrive on the 5G network. Therefore, all packets in a TSN flow experienced the same latency. This is shown in Fig. 7.b, which shows the average standard deviation experienced by the different TSN flows with the proposed and reference scheme. Fig. 7.b shows that the average standard deviation experienced with the reference scheme is zero. On the other hand, the proposed scheduling scheme allocates radio resources individually for each packet of a TSN flow within a time window of size *HP* (the resources are allocated periodically with period *HP*). Thanks to this flexibility, the proposal assigns for each packet of a TSN flow the radio resources that minimize the latency based on the availability of unassigned resources at each moment. This results in a reduction of the latency experienced in the 5G network at the expense of an increase in the standard deviation, as shown in Fig. 7.b. The standard deviation will be minimized by the TSN translators that will hold and forward the packets at the departure time indicated by the TSN network. It is essential to highlight that the important thing is that packets arrive at the egress port of the 5G logical bridge (the TSN translator) at the time required to be transmitted to the next node in the path (a TSN bridge or the end device).

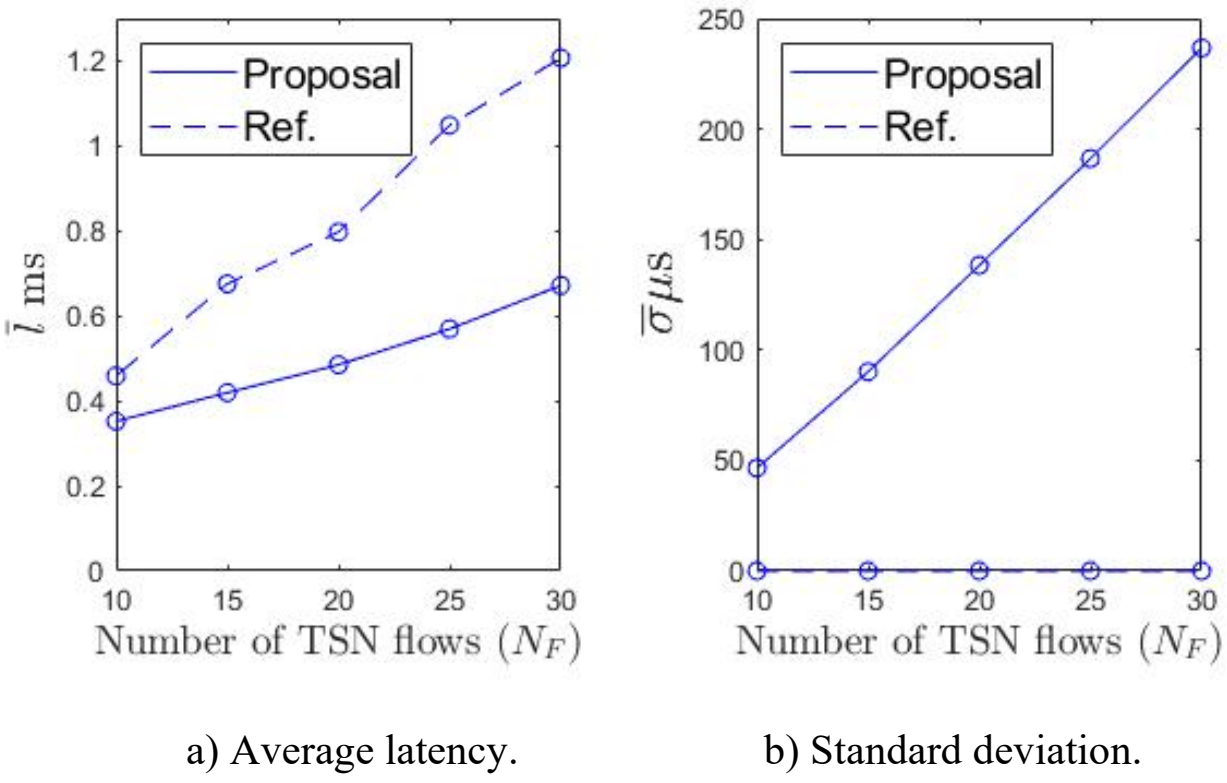

a) Average latency. b) Standard deviation.

Fig. 7. Average latency and standard deviation of the proposed and reference scheduling schemes as a function of the number of TSN flows when $l_{e2e_i}$ is equal to the periodicity.

## VI. Conclusions

This paper has presented and evaluated a novel Configured Grant scheduling scheme for 5G networks integrated with TSN networks. This paper considers the 5G-TSN integration model where 5G is integrated into the TSN network as a bridge. The proposed scheduling scheme aims to coordinate its scheduling decision with the scheduling performed in TSN to efficiently and effectively meet the end-to-end latency and determinism requirements of TSN traffic. To this end, the proposed scheduling scheme uses the information provided by TSN about the characteristics of the TSN traffic (periodicity, packet size, etc.) to satisfy the maximum latency that based on the TSN scheduling decision must be experienced in the 5G network. Common CG scheduling schemes allocate radio resources periodically for each TSN flow based on the periodicity of the packets. This may result in the use of the same radio resource for the transmission of some packets by different TSN flows due to the different periodicity of the packets of different TSN flows. The proposed scheduling scheme solves this issue by configuring several grants for each TSN flow in order to avoid radio resource allocation conflicts among different TSN flows. The results achieved in this paper have demonstrated that the proposed scheduling scheme considerably increases the number of TSN flows that can be satisfactorily served in the integrated 5G-TSN network compared to common CG schemes. Furthermore, the average latency experienced in the 5G network is reduced with the proposed scheme.